\documentclass[aps,twocolumn,showpacs,amsmath,amssymb,prl,floatfix]{revtex4-1}

\usepackage{graphicx,float}
\usepackage{bm}
\usepackage{mathrsfs}

\usepackage{color}

\allowdisplaybreaks[2]

\begin{document}

\title{Collectivity-assisted ground state cooling of a nanomechanical resonator}

\author{Keyu \surname{Xia}}

\author{J\"{o}rg \surname{Evers}}%
\affiliation{Max-Planck-Institut f\"{u}r Kernphysik, Saupfercheckweg 1, D-69117 Heidelberg, Germany}

\date{\today}

\begin{abstract}
We discuss cooling of a nanomechanical resonator to its mechanical ground state by coupling it to a collective system of two interacting flux qubits. We find that the collectivity crucially improves  cooling by two mechanisms. First, cooling transitions proceed via sub-radiant Dicke states, and the reduced line width of these sub-radiant states suppresses both the scattering and the environmental contribution to the final phonon number. Second, detrimental carrier excitations without change in the motion of the resonator are suppressed by collective energy shifts.
\end{abstract}



\maketitle
Micro- and nanomechanical resonators (NAMRs) are currently in the focus of interest because of the fascinating applications they promise~\cite{PhysToday58p36}.
Access to these applications requires cooling of the mechanical motion to the quantum mechanical ground state, which as yet is an unsolved task. Different routes towards the ground state have been suggested. For example, backaction cooling or sideband cooling in optomechanical systems has been proposed and demonstrated via  cavity-assisted radiation pressure~\cite{NaturePhys5p485,*NaturePhys5p509,*NaturePhys5p489,Science321p1172,PRL99p073601, Nature444p71,PRL99p093901,*PRL99p093902}.
In the electromechanical domain, cooling methods have been suggested which are based on auxiliary devices such as superconducting transmission lines~\cite{PRL101p197203}, superconducting single-electron transistors~\cite{Nature443p193}, or quantum dots~\cite{PRL92p075507,PRB79p075304}.
Just as in laser cooling of atoms or ions, backaction cooling or sideband cooling can only reach the ground state if the mechanical resonance frequency exceeds the bandwidth of the cavity or auxiliary quantum system. This so-called strong confinement regime (SCR) restricts the accessible parameter range, and is often difficult to implement. Active feedback cooling~\cite{PRL99p073601,Nature444p75,PRL80p688} is a candidate for overcoming this limit. However, it typically requires difficult control and precise measurement of displacement of the resonator.
Another route is to make use of quantum interference. In~\cite{PRL102p207209,*PRL103p223901}, a system was proposed in which the mechanical displacement of the oscillator changes the damping, rather than the frequency of a coupled cavity. This leads to destructive interference in the noise contributions, and thus to the possibility of ground state cooling already outside the SCR. Recently, we analyzed a cooling scheme for NAMR based on electromagnetically induced transparency (EIT)~\cite{PRL103p227203}. EIT cooling has originally been proposed~\cite{PRL85p4458} and demonstrated~\cite{PRL85p5547} for trapped ions, and works by eliminating unwanted heating transitions via destructive quantum interference.

\begin{figure}[b]
\centering
\includegraphics[height=4cm,width=8cm]{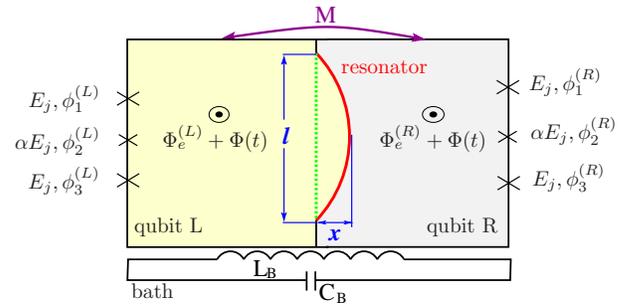}
\caption{\label{fig:system}(Color online) A nanomechanical resonator coupled to two flux qubits. The flux qubits interact with each other through their mutual inductance $M$ and are damped via a common bath modeled as a $LC$ circuit.}
\end{figure}

In this Letter, we propose an efficient ground state cooling scheme for a NAMR without counterpart in the cooling of atoms or ions. The NAMR is coupled to two interacting flux qubits as shown in Fig.~\ref{fig:system}. The mutual interaction of the qubits gives rise to  frequency-shifted collective qubit states with modified decay rates. We find that in a suitable cooling field configuration, the collective frequency shift effectively eliminates detrimental qubit carrier excitations without change in the NAMR motion. At the same time, cooling via the sub-radiant collective qubit state suppresses both contributions from the thermal environment and the scattering of the cooling field to the cooling limit.
Due to these collective effects, our scheme offers efficient ground state cooling already for rather small driving fields, and for NAMR operating outside the strong-confinement  regime. An implementation is further assisted by the simple level structure of the qubit  part compared, e.g., to EIT cooling, as the flux qubits are modelled as two-level systems operating close to the optimum point. Unlike in backaction cooling and feedback cooling, the final NAMR state has no coherent shift in its phonon number.

\begin{figure}[t]
\centering
\includegraphics[width=\linewidth]{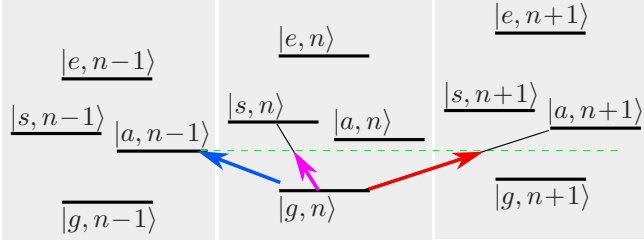}
\caption{(Color online) Energy level diagram for the NAMR coupled to two interacting flux qubits. Cooling (heating) transitions $|n\rangle\to |n\pm1\rangle$ proceed via the narrow collective sub-radiant state $|a\rangle$ and are denoted by blue (red) arrows. Carrier excitation $|n\rangle\to |n\rangle$ (purple arrow) is possible only via the strongly detuned symmetric collective state. }\label{fig:Lvldiag}
\end{figure}

Before we proceed with the analysis, we start by explaining the physical mechanisms behind collectivity-assisted cooling. First, we consider the two interacting qubits without coupling to the motional state $|n\rangle$ of the oscillator~\cite{PRL94p090501}. Qubit $j$ ($j\in\{L,R\}$) can be visualized as a two-level quantum system with ground state $|g_j\rangle$, excited state $|e_j\rangle$, and decay rate $\gamma$. The two qubits $L$ and $R$ are coupled via their mutual inductances, such that the system dynamics can conveniently be described in the collective-state basis
%
$|g\rangle=|g_Lg_R\rangle$, 
$|s\rangle=(|e_Lg_R\rangle+|g_Le_R\rangle )/\sqrt{2}$,
$|e\rangle=|e_Le_R\rangle$, and
$|a\rangle=(|e_Lg_R\rangle-|g_Le_R\rangle)/\sqrt{2}$.
%
In a strongly interacting system~\cite{FicekQIC,PRB76p100505}, the anti-symmetric state $|a\rangle$ is characterized by a significantly reduced decay rate $\Gamma_a\ll \gamma$, while the symmetric state has an enhanced rate $\Gamma_s \approx 2 \gamma$. At the same time, the two  states are shifted in energy due to the qubit-qubit interactions. Analogously, a driving field which is applied symmetrically to the two qubits will 
couple to transitions $|g,n\rangle \leftrightarrow |s,n\rangle$ and $|s,n\rangle \leftrightarrow |e,n\rangle$, but not to those via $|a,n\rangle$.
Next, we consider in addition the coupling to the motional eigenstates $|n\rangle$ of the NAMR. A relevant part of the total energy level spectrum is shown in Fig.~\ref{fig:Lvldiag}. It turns out that the NAMR motion couples anti-symmetrically to the two qubits, because a movement of the oscillator always increases the loop area of one of the qubits, while decreasing the area of the other. Therefore, the symmetrically applied driving field still couples the ground state $|g,n\rangle$ to the symmetric state $|s,n\rangle$ without change in the motion,  but it couples $|g,n\rangle$ to the anti-symmetric states $|a,n\!\pm 1\!\rangle$ if the motional state increases or decreases. The three leading excitation channels from the ground state are depicted as arrows in Fig.~\ref{fig:Lvldiag} for a field in resonance with the red sideband to state $|a, n\!-\!1\rangle$.
Then, as in sideband cooling~\cite{PRL92p075507}, this leads to a cooling of the NAMR, but with substantially inproved performance due to the scattering via collective states. The cooling limit crucially depends on the ratio of the NAMR eigenfrequency $\nu$ to the width of the scattering state. Sideband  cooling to the motional ground state is only possible in the resolved regime in which $\nu$ exceeds the scattering width. This explains the first advantage of collectivity-assisted cooling: The cooling limit is determined by $\Gamma_a/\nu$ rather than $\gamma/\nu$ as in most other cooling schemes. In particular, in the common case $\Gamma_a \ll \nu < \gamma$, a single qubit would be in the non-resolved regime, whereas the collective system effectively becomes resolved.
The second advantage is a suppression of carrier excitations $|n\rangle \to |n\rangle$, which also leads to a heating. This suppression is achieved mainly by the energy shift of the collective states $|s\rangle$ and $|a\rangle$, which moves the symmetric carrier scattering channel out of resonance with the cooling field.
Thirdly, the narrow width $\Gamma_a$ also suppresses the non-resonant  heating processes $|n\rangle \to |n+1\rangle$ as compared to the single-qubit case. %
These advantages together lead to a greatly improved collectivity-assisted cooling performance.


We now proceed with a quantitative analysis.
The NAMR has effective mass $M_{\rm eff}$, length $l$, frequency $\nu$, and quality factor $Q$. It can be treated as a harmonic quantum resonator with Hamiltonian $H_r=\hbar\nu b^\dag b$. The quantized displacement operator is $x=X_0(b+b^\dag)$, where $b$ is the annihilation operator. $X_0=\sqrt{\hbar/2M_{\rm eff}\nu}$ denotes the zero point fluctuation of the NAMR.
%
Each flux qubit consists of superconducting loops with three Josephson junctions, two identical and one junction smaller by a factor $\alpha$. The qubits are exposed both to a constant magnetic field $\textbf{B}$ perpendicular to the plane and to a common driving microwave field $\Phi(t)$, which induces a time-dependent magnetic flux (TDMF).
Following a standard treatment of the coupled flux qubits~\cite{PRL96p067003}, both qubits are modeled as two-level systems. The free system Hamiltonian of flux qubits and NAMR is given by $H_0=\hbar \nu b^\dag b+\hbar \Delta (R_{ee}-R_{gg})+\hbar \Lambda (R_{ss}-R_{aa})$. The qubit operators are defined as $R_{jk}=|j\rangle\langle k|$  $(j,k\in\{e,s,a,g\})$. The detuning $\Delta=\omega_0-\omega_L$ is the difference of the qubit transition frequency $\omega_0$ and the TDMF frequency $\omega_L$. $\Lambda$ includes the always-on coupling and the collective shifts induced by the coupling of the two qubits~\cite{PRL96p067003,PRB79p024519}. The qubits operate near their optimal point $f=0.5$, and are assumed degenerate, which is reasonable if their transition energy difference is much smaller than $\Lambda$. We assume that the qubit transition frequencies are large enough to neglect thermal excitations.

The qubit-NAMR interaction is modelled by 
$H_I=\sqrt{2}\hbar \Omega(R_{es}+R_{sg}+R_{se}+R_{gs})+\sqrt{2}\hbar \Omega \eta (b+b^\dag) (R_{ag}+R_{ga}-R_{ea}-R_{ae})$.
The Rabi frequencies are 
 $\hbar\Omega_j=\frac{\alpha E_J}{2}\frac{2\pi}{\Phi_0}A \mathcal S_j$
 with coupling energy $\alpha E_J$, $\hbar\Omega_L=\hbar\Omega_R=\hbar\Omega$, $\mathcal S_j = \langle e_j|\sin(2\varphi_p^{(j)} + 2\pi f) |g_j\rangle$ and $\mathcal C_j = \langle e_j|\cos(2\varphi_p^{(j)} + 2\pi f) |g_j\rangle$. The phase  $\varphi_p^{(j)}=(\varphi_1^{(j)}-\varphi_2^{(j)})/2$. A contribution from the momentum can be neglected. The Lamb-Dicke parameters evaluate to $\eta_j=\varsigma_j BlX_0 2\pi / \Phi_0$, where $\varsigma_j=\mathcal C_j / \mathcal S_j$. We assume that $\eta = \eta_L=\eta_R$.
%
Our system couples to an environment of temperature $T$, such that the thermal NAMR occupation is $N_i=\left[\exp(\hbar\nu/k_B T)-1\right]^{-1}$. In Born-Markov, rotating wave, and Lamb-Dicke approximations, the master equation is
\begin{subequations}
\label{eq:MEq}
\begin{align}
 \dot \rho =&-\frac{i}{\hbar}[H_{0}+H_I,\rho]
+ \mathscr{L}_{SE} \rho  +\mathscr{L}_\phi \rho\nonumber \\
&+[N_i+1]\mathscr{L} \left(\nu/Q,b\right)+N_i\mathscr{L} \left(\nu/Q,b^\dag\right)\,,
\label{sub-master}\\
\mathscr{L}_\phi=&\sum_j {\mathscr{L}\left(\Gamma_\phi/2,|e_j\rangle \langle e_j|-|g_j\rangle \langle g_j|\right)\rho}\,,\\
\mathscr{L}_{SE}\rho =&
    \mathscr{L}(\Gamma_s,R_{se}+R_{gs})  +  \mathscr{L}(\Gamma_a,R_{ae}-R_{ga})
 \nonumber \\
  +& \mathscr{L}\left(\Gamma_s\eta^2,(R_{ae}-R_{ga})(b+b^\dag)\right) \nonumber\\
  +& \mathscr{L}\left(\Gamma_a\eta^2,(R_{se}+R_{gs})(b+b^\dag)\right) \nonumber\\
 +& \mathscr{L}(\Gamma_s\eta, (R_{ga}-R_{ae})(b+b^\dag), R_{se}+R_{gs})\nonumber\\
+&\mathscr{L}(\Gamma_a\eta, (R_{se}+R_{gs})(b+b^\dag), R_{ga}-R_{ae})\,,\\
\mathscr{L} (\kappa,A) &\rho =\frac{\kappa}{2}\left\{2A\rho A^\dag- [A^\dag A, \rho]_+  \right\} \,,  \\
\mathscr{L} (\kappa,A_1,&A_2) \rho =\frac{\kappa}{2}\sum_{i\neq j}\left\{2A_i\rho A_j^\dag- [A_j^\dag A_i, \rho]_+  \right\} \,,
\end{align}
\end{subequations}
with the damping coefficients $\Gamma_{s}=\gamma+\gamma_{12}$ and $\Gamma_a=\gamma-\gamma_{12}$ and $\kappa=\nu/Q$. Here, $\gamma$ and $\gamma_{12}$ are the Einstein $A$ coefficient and the dipole-dipole cross damping rate, respectively. $\Gamma_\phi$ is the pure dephasing of a single qubit.
In Eqs.~(\ref{eq:MEq}), $\mathscr{L}_{SE}\rho$ describes the spontaneous emission of the flux qubits. The additional pure dephasing in described by  $\mathscr{L}_\phi \rho$. The last two terms in Eq.~(\ref{sub-master}) result from the coupling of NAMR and thermal environment.

\begin{figure}[t]
\centering
\includegraphics[width=\columnwidth]{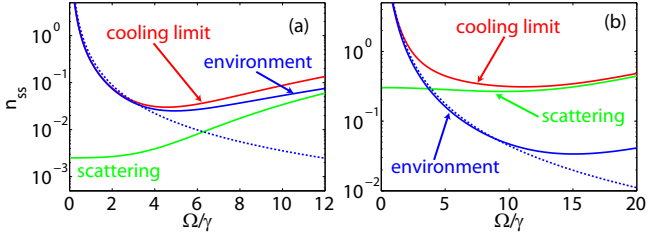}
\caption{\label{fig:nssfld}(Color online) Final average phonon number $n_{\rm ss}$ and its contributions as a function of the cooling field Rabi frequency $\Omega$. Parameters are $N_i=400$, $Q=10^6$, $\Gamma_a=0.1\gamma$,and $\eta=3\times 10^{-3}$. Dephasing is  (a) $\Gamma_\phi=0$ and (b) $\Gamma_\phi=0.5\gamma$. A fit to the environmental contribution by Eq.~(\ref{eq:nss}) is shown as dashed line, with (a) $C=0.16$ and (b) $C=2$.}
\end{figure}
%


As we start cooling from thermal equilibrium giving an initial phonon number $N_i=N(\nu)$, the final phonon number $n_{\rm ss}$  is determined by two contributions: one proportional to the initial phonon number and the other from the scattering of the cooling field,
\begin{equation}\label{eq:nss}
n_{\rm ss}=C(\eta,\Omega,\Gamma_a)\Gamma_a\nu\frac{N_i}{(\eta \Omega)^2Q}+G(\eta,\Omega,\Gamma_a) \left(\frac{\Gamma_a}{4\nu}\right)^2 \,.
\end{equation}
 Here we introduce coefficients $C(\eta,\Omega,\Gamma_a)$ and $G(\eta,\Omega,\Gamma_a)$ to correct the deviation of our approximate analytical expression from the numerical results. 
The analytical parts of Eq.~(\ref{eq:nss}) allow for a comparison with other cooling methods. We find that the environmental contribution to the cooling limit is suppressed by a factor of $C\Gamma_a\nu$ in comparison with  sideband cooling \cite{PRL92p075507} and backaction cooling \cite{PRL99p093901,PRL99p093902}. For this comparison, we assume effective Lamb-Dicke parameters defined as $\eta_{\rm eff}=\eta \Omega/\gamma$ in our case and $\eta_{\rm eff}=\eta \sqrt{n_{\rm max}}$ in the backaction cooling scheme, where $\eta=(d\omega_c/dx)(X_0/\nu)$ with the cavity  resonance frequency $\omega_c$ and $n_{\rm max}$ as the photon number in resonance \cite{PRL99p093901}.  Interestingly, as the environmental contribution is proportional to $\nu$, the nonresolved regime with small $\nu$ is favourable.  We also find that in our cooling scheme there is no coherent shift in the final phonon number, other than in backaction cooling~\cite{PRL99p093901} and feedback cooling~\cite{PRL80p688}. This is crucial for many applications such as in quantum information science~\cite{PRL93p190402,*PRL91p130401}.

\begin{figure}[t]
\centering
\includegraphics[width=\columnwidth]{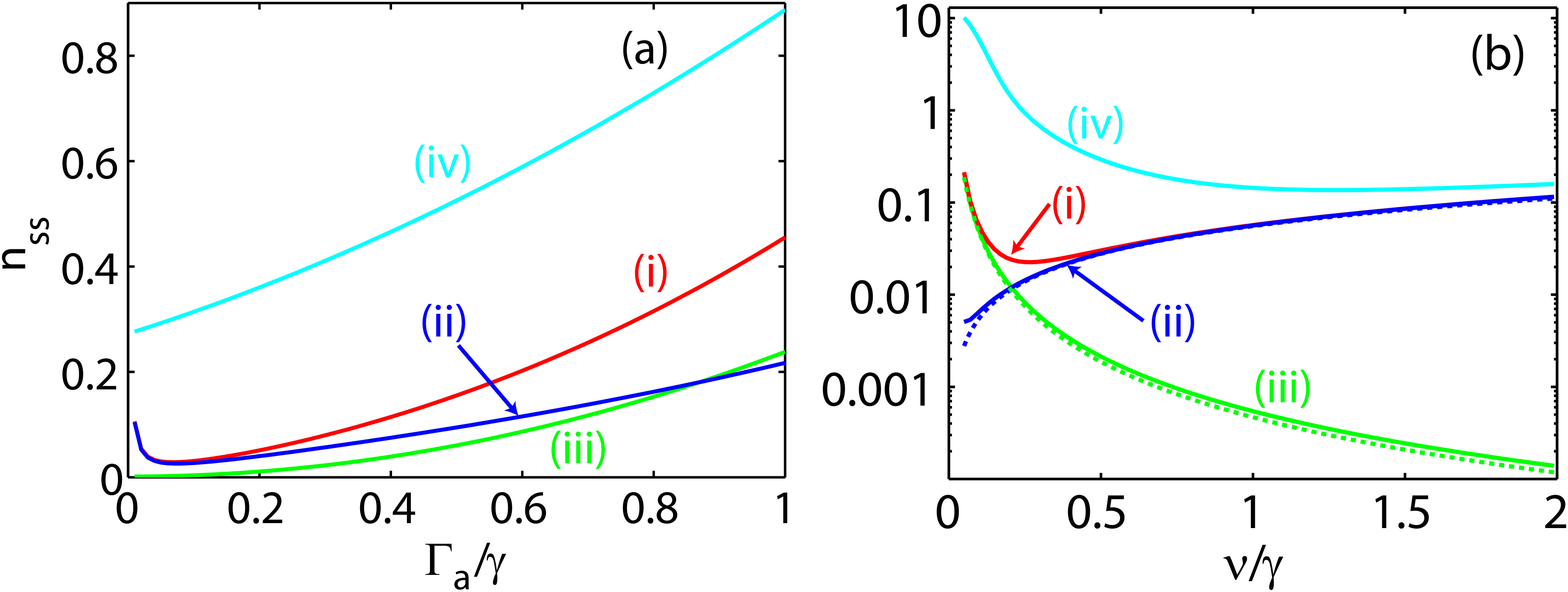}
\caption{\label{fig:4}(Color online)
The final resonator phonon number $n_{\rm ss}$ [curves (i) and (iv)] and its contributions from environment (ii) and scattering (iii).
The parameters are $\Gamma_\phi=0$ and $\Omega=4\gamma$ in curves (i)-(iii), and $\Gamma_\phi=0.5\gamma$ and $\Omega=10\gamma$ in (iv). Further, $\eta=3\times 10^{-3}$, and $Q=10^6$. Dashed lines show analytical fits to the numerical data using Eq.~(\ref{eq:nss}).
(a) shows the results versus the sub-radiant decay width $\Gamma_a$ with $\nu=0.5\gamma$, and (b) against the resonator frequency $\nu$ with $\Gamma_a = 0.05\gamma$.}
\end{figure}

We now turn to a full numerical study of Eqs.~(\ref{eq:MEq}).
Typical flux qubit parameters we envisage are $\alpha=0.8, E_J/E_C=100,E_J=1000~$GHz$~(I_C\sim300$~nA), where $E_J$ and $E_C$ are the Josephson and charging energies of junction, respectively. We assume a bias flux $f_b=0.5005$. We calculate transition frequency $\omega_0/E_J\sim 4.5\times 10^{-3}$, and transition matrix element  
$\mathcal S_j \sim -0.16$ and $\varsigma\sim 0.05$. We assume a decay rate of about $2\pi\times 2$~MHz, and experimental results show that this rate can remain similar over a wide range of bias \cite{PRL97p167001}. Pure dephasing is assumed as half of the decay rate for a typical noise spectral density $(\mu \Phi_0)^2$/Hz at $f_b=0.5005$ \cite{PRL97p167001}. Note that the qubit parameters can be engineered over a wide range~\cite{PhysToday58p42,*PRB60p15398,*Science314p1589,*QIP8p261}.
Regarding the resonator, a double clamped NAMR with size of $300~\mu$m$\times 100~$nm$\times 100~$nm made from silicon nitride can be manufactured  with frequency $\nu \approx 2\pi\times 1~$MHz and quality $Q\approx10^6$ \cite{APL92p013112}. This NAMR leads to a Lamb-Dicke parameter $\eta\sim 0.003$ for $B\lesssim 2T$. Thus our system can be realized based on NbN superconducting qubits. 
Our sample chip could be placed in a dilution refrigerator of $20~$mK giving rise to a initial phonon number $N_i\sim 400$. Assuming that the qubits is driven by a $100\Omega$ microwave line coupled by mutual inductance $\sim 50~$fH, a Rabi frequency of $10\gamma$ only requires  $\sim 0.5~$nW input power. This small input power helps to reduce nonlinear and phase noise \cite{PRL101p197203}.
Throughout our investigation, we fix $\Lambda=500\gamma$ and apply the TDMF with detuning $\Delta=\nu+\Lambda$ corresponding to a red sideband excitation $|g,n\rangle\to |a,n\!-\!1\rangle$. Note that a large mutual inductance  $M$ can lead to couplings $\Lambda$ up to  a few GHz~\cite{PRL94p090501}.

%
An example for the dependence of the final phonon number $n_{\rm ss}$  on the TDMF strength $\Omega$ is shown in Fig.~\ref{fig:nssfld}. Without pure dephasing,  one can ideally cool a NAMR from $N_i=400$ to a final phonon number $n_{\rm ss}$ smaller than $0.05$ in the range of $3 \leq \Omega/\gamma \leq 8$, see Fig.~\ref{fig:nssfld}(a).
This corresponds to a ratio $N_i / N_f$ of initial to final phonon number exceeding $10^3$.
By fitting the environmental contribution, we find that $C=0.16$. Our analytic formula Eq.~(\ref{eq:nss}) becomes invalid for $\Omega$ exceeding $4\gamma$.
Taking into account an additional pure dephasing $\Gamma_\phi=0.5\gamma$, we find that  $n_{\rm ss}$ is still lower than $0.5$ in the range of $4 \leq \Omega/\gamma \leq 20$, see Fig.~\ref{fig:nssfld}(b). At $\Omega=10\gamma$, the ground state occupation is $76\%$. Further investigation show that a ground state cooling is possible as long as $\Gamma_\phi<\gamma$.

According to  Eq.~(\ref{eq:nss}), both the environmental and the scattering contributions to the cooling limit are suppressed by the reduced decay rate $\Gamma_a$ of the sub-radiant state. This result is confirmed by our numerical results shown in Fig.~\ref{fig:4}(a).
Interestingly, an optimal cooling limit is obtained for $\Gamma_a\approx 0.05\gamma$. In this point, the total final phonon number is only $0.03$, which corresponds to a $97\%$ occupation of ground state. Further decrease of $\Gamma_a$ leads to an unexpected rapid increase of the cooling limit. The origin of this increase is that due to level shifts, the qubit transition frequency is slightly off-resonant with the driving field frequency. If the scattering state becomes too narrow, then the driving field cannot excite the cooling transitions any more. We verified this interpretation by a suitable slight change in the driving field frequency, which allows to achieve effective cooling also for smaller $\Gamma_a$.
With additional pure dephasing $\Gamma_\phi=0.5\gamma$, the optimum cooling limit is obtained for $\Omega=10\gamma$. For $\Gamma_a=0.05\gamma$, the steady state is  $n_{\rm ss}\approx0.29$, which indicates $78\%$ occupation of ground state. 

Next, we study the dependence of the cooling limit on the  resonance frequency $\nu$ of the NAMR, see Fig.~\ref{fig:4}(b). The environmental contribution to $n_{\rm ss}$ is proportional to $\nu$. Our scheme thus provides particularly efficient cooling of NAMRs with low frequencies. But since the scattering contribution is inversely proportional to the square of $\nu$, this part becomes eventually dominant with decreasing $\nu$. We find an optimal point of $\nu\approx 0.25\gamma$ where a minimum final phonon number of $0.02$ is achieved. This number agrees well with the optimal value $\nu_{opt}=\sqrt[3]{\eta^2\Omega^2 Q\Gamma_a G/8CN_i}$ evaluated from Eq.~(\ref{eq:nss}) with $C=0.4$ and $G\sim 3$. When including a pure dephasing of $\Gamma_\phi=0.5\gamma$, an optimal cooling limit of $n_{\rm ss}=0.15$ is obtained for $\Omega=10\gamma$ and $\nu=\gamma$.

In summary, we have presented an efficient ground state cooling scheme for a NAMR which operates by coupling the NAMR to two interacting flux qubits. We found that collectivity enhances the cooling performance in two crucial ways. First, the collective frequency shift of the qubit states effectively eliminates unwanted carrier excitations. Second, cooling via the sub-radiant Dicke state suppresses both contributions from the thermal environment and the scattering of the cooling field to the cooling limit. As a result of this, ground state cooling can be achieved  already in the non-resolved regime, and using rather small driving fields.

%

\end{document}